\documentclass[11pt]{article}
\usepackage{amssymb,cite}
\usepackage{graphicx}
\usepackage{subfigure}
\usepackage[subfigure]{ccaption}
\usepackage{latexsym}
\usepackage{epsfig}

\def\thefootnote{\fnsymbol{footnote}}

\newcommand{\eq}{\begin{equation}}
\newcommand{\en}{\end{equation}}
\newcommand{\be}{\begin{equation}}
\newcommand{\ee}{\end{equation}}
\newcommand{\eqa}{\begin{eqnarray}}
\newcommand{\ena}{\end{eqnarray}}
\newcommand{\ba}{\begin{eqnarray}}
\newcommand{\ea}{\end{eqnarray}}

\newcommand{\ZZ}{\hbox{{\rm Z{\hbox to 3pt{\hss\rm Z}}}}}

\begin{document}
\begin{titlepage}
\vskip0.5cm
\begin{flushright}
DFTT 4/07\\
\end{flushright}
\vskip0.5cm
\begin{center}
{\Large\bf Universal power law behaviors in genomic sequences and evolutionary models}
\end{center}
\vskip1.3cm
\centerline{Loredana~Martignetti and
Michele~Caselle}
 \vskip1.0cm
 \centerline{\sl  Dipartimento di Fisica
 Teorica dell'Universit\`a di Torino and I.N.F.N.,}
 \centerline{\sl Via Pietro~Giuria 1, I-10125 Torino, Italy}
 \centerline{\sl
e--mail: \hskip 1cm
 (caselle)(martigne)@to.infn.it}
 \vskip0.4 cm
\begin{abstract}
We study the length distribution of a particular class of DNA sequences known as 5'UTR exons. These exons belong
to the messanger RNA of protein coding genes, but they are not coding (they are located upstream of the coding portion of the mRNA) and
are thus less constrained from an evolutionary point of view. We show that both in mouse and in human 
these exons
show a very clean power law decay in their length distribution and suggest a simple evolutionary model which may explain this
finding. We conjecture that this power law behaviour could indeed be a general feature of higher eukaryotes.

\end{abstract}
\end{titlepage}

\setcounter{footnote}{0}
\def\thefootnote{\arabic{footnote}}
\section{Introduction}
In these last years lot of efforts have been devoted in trying to find universal laws in nucleotide
distributions in DNA sequences. A typical example was the identification more than ten years ago
of long range correlations in the base composition  of DNA (see for instance~\cite{Li:1997,Arneodo:1995} and 
references therein). 
With the availability of complete sequenced genomes, the correlation property of length sequences has been studied separately for
coding and non coding segments of complete bacterial genomes, showing a rich variety of behaviour for different kinds of sequences
 \cite{yu:2000,yu:2001}.
This line of research has been recently
extended to the search of similar universal distribution for more complex features of eukaryotic DNA sequences
like for instance 5' untranslated regions (UTR) lengths~\cite{Lynch:2005}, UTR introns~\cite{Hong:2006} or strand asymmetries in
nucleotide content~\cite{Brodie:2005,Touchon:2004}. The main reason of interest for this type of analyses is the
search of general rules behind the observed universal behaviours. The hope is to get in this way
new insight in the evolutionary mechanisms shaping higher eukaryotes genomes and to understand
functional role of the various portions of the genome. An intermediate important step of this process is the
construction of  simplified (and possibly exactly solvable)
stochastic models to describe the observed behaviours. This is the case for instance of the  model 
discussed in~\cite{Messer:2005} for base pair correlations or the model proposed in~\cite{Lynch:2005} 
for the 5'UTR length.
In this paper we describe a similar universal law for the exon length in the 5'UTR of the
human and mouse genomes. Looking at the 5'UTR exons collected in the existing genome databases for the two
organisms we shall first show that they follow with a high degree of confidence a power law 
distribution with a decay
exponent of about $2.5$ and then suggest a simple solvable model to describe this behaviour.

We shall also compare the impressive stability of the power law decay of 5'UTR exons with the distributions in
the case of the 3'UTR and coding exons which turn out to be completely different. This is most probably due to
the different evolutionary pressures to which are subject the three types of sequences.

We think that the behaviour that we observed  should indeed be a general feature of higher eukaryotes, 
however its
identification requires a very careful annotation of 5'UTR regions which exist for the moment only for human and
mouse (see tab. below).

This paper is organized as follows. After a short introduction to the biological aspects of the problem (sect. 2) 
we discuss the exon length distribution in sect.3 . Sect.4 is then devoted to the discussion of a simple
stochastic model which gives as equilibrium distribution the observed power like behaviour. 
Details on the model are collected in the Appendix.

\section{Biological background}

In eukaryotic organisms, DNA information stored in genes is translated into proteins through a series of complex
 processes, carefully controlled at each step by specific regulatory mechanisms activated by the cell. 
 In particular, two crucial events in this process are the production of an intermediate molecule, the messanger 
 RNA (mRNA) transcript, and the translation of the mRNA into proteins. 
The cell provides fine regulatory systems to regulate the gene expression both at transcriptional and 
post-transcriptional level, using several cis--acting signals located in the DNA sequence. A common molecular 
basis for much of the control of gene expression (whether it occurs at the level of initiation of transcription, 
mRNA processing, translation or mRNA transport) is the binding of protein factors and specific RNA elements to 
regulatory nucleic acid sequences.

Once mRNA is transcribed, it usually contains not only the protein coding sequence, but also additional segments,
 which are transcribed but not translated, namely a flanking  5' untranslated region (5' UTR) and a final 
 3' UTR \footnote{5' and 3' refer to the position (5' and 3' respectively) of the
 carbon atoms of the mRNA backbone at the two extrema of the mRNA and are conventionally used to denote the
 ``upstream'' (5') and "downstream" (3') sides of the mRNA chain.}. 
 Nucleotide patterns or motifs located in 5' UTRs and 3' UTRs are known 
 to play crucial roles in the post-transcriptional regulation. Most of the  primary transcripts
 of euKaryotic genes also contain sequences (named ``introns'') which are eliminated during  
 a maturation process named ``splicing''. The sequences which survive this splicing process are named "exons"
 they are glued together by the splicing machinery and form the mature mRNA transcript. Both the UTRs and
 the coding portions of the mRNA are usually composed by the union of several exons. It is thus possible to classify
 the exons as coding, 3'UTR and 5'UTR depending on the portion of the mRNA to which they
 belong\footnote{Obviously in several cases one can have exons which are partially included in one of the two UTR
 regions and partially in the coding portion of the mRNA. These mixed exons were excluded from our analysis.}
 
  A cell can splice the 
``primary transcript" in different ways and thereby make different polypeptide chains from the same gene 
(a process called alternative RNA splicing) and a substantial proportion of
higher eukaryotic genes 
(at least a third of human genes, it is estimated) produce multiple proteins in this way (isoforms), 
thanks to special signals in primary mRNA transcripts.

Some hints about the 5' and 3' role in gene expression can be derived from a quantitative analysis of 
UTR length. 

Recent large scale databases suggest that the mean 3' UTR length in human transcript is nearly four times longer 
than the mean human 5'UTR length \cite{Pesole:2002} and that the evolutionary expansion  of 3'UTR in higher 
vertebrates, not observed in 5' UTR, is associated to their peculiar regulatory role. Very recent works revealed 
the existence of an extremely important post-transcriptional regulatory mechanism, performed by an abundant 
class of small non coding RNA, known as  microRNA (miRNA), that recognize and bind to multiple copies of 
partially complementary sites in 3'UTR of target transcripts, without involving 5'UTR \cite{Bartel:2004,Hannon:2004,Rajewsky:2006}.

Differently, 5'UTR sequences are expected to be constrained mainly by splicing process and  translation 
efficiency.
The exons in the 5' UTR regions are usually termed ``non coding exons'', since they are not included in 
the protein coding portion of the transcript. However, their characteristics, as their length, secondary 
structure and the presence of AUG triplets upstream of the true translation start in mRNA, known as 
upstream AUGs, have been shown to affect the efficiency of translation and to be preserved in the evolution
 of these sequences \cite{Pesole:2005,Koonin:2005,Lynch:2005}. 5'UTR exons length can vary between few tens 
 until hundreds of nucleotides, without typical length scale around favourite size, and the lower and upper 
 bounds of this distribution is likely to be shaped by splicing and translation efficiency: exons that are 
 too short (under 50 bp) leave no room for the spliceosomes (enzymes that perform the splicing) to operate 
 \cite{sorek}, while exons that are too long can contain signals that affect translation efficiency. 5'UTR 
 ``non coding exons'' are also free from selective pressure acting on coding exons, which strongly preserves 
 the amminoacid information written in triplets of nucleotides in the protein coding exons. 

For these reasons, in our analysis we decided to construct strictly disjoint subsets of exons, according to 
their position in the transcript (5'UTR exons, protein coding exons or 3'UTR 
exons)\footnote{Obviously in several cases one can have exons which are partially included in one of the two UTR
 regions and partially in the coding portion of the mRNA. These mixed exons were excluded from our analysis.}. 
 Moreover, we created 
a non redudant genome-wide datasets of exons, considering only one isoform for each gene, the most extended one. 

Curated information about DNA sequences and annotation of eukaryotic organisms are provided by the Ensembl 
project, based on a software system which produces and maintains automatic annotation on selected eukaryotic 
genomes \cite{ensembl:2007}.
\section{Analysis of exons distribution}
We downloaded from the Ensembl 
database (release 40 \cite{ensembl:2007}) all the available transcripts annotated as protein coding
for different organisms, and we created a filtered dataset of non redundant exons, considering 
the most extended transcript for each gene. We eliminated all the exons with mixed annotations and grouped the
remaining ones in three classes: 5'UTR, protein coding exons, and 3'UTR. 

Plotting the length distribution of exons, separately for 
5'UTR, coding exons and 3'UTR, we clearly observe different behaviours, which we think should reflect 
different evolutionary 
constrains acting on these classes 
of DNA sequences (Fig.1 a,b,c). In particular, the 5' UTR exons size distribution  shows a remarkably 
smooth power decay for large enough values of the exon length.
To assess this point and to evaluate the threshold above which the power law behaviour starts,
we fitted the observed distributions with a power law:
\begin{equation}
N(l)=l^{-\alpha}
\end{equation}
where $N(l)$ is the number of exons of length $l$.

In order to evaluate the goodness of the fits that we performed, we divided the set of all exons into 18
equivalent bins and then assumed the variance of these bins as an indication of the statistical uncertainty of
our estimates (results are independent from the binning choice). This allowed us to perform a meaningful
 $\chi^2$ test on the fits. This test is commonly used when an assumed distribution is evaluated against the 
 observed data \cite{statistics}. The quantity $\chi^2$ may be thought of as a measure of the discrepancy 
 between the observed values and the respective expected values. It is convenient to compute the reduced 
 chi square $\widetilde{\chi}^2$ (i.e. the ratio $\chi^2/(N_p-N_f)$ where $N_p$ is the number of points included in the
 fit and $N_f$ the number of parameters of the fit). With this normalization one can immediately see if the
 fitting function correctly describes the data (which requires $\widetilde{\chi}^2 \leq 1$). When 
 instead  $\widetilde{\chi}^2 > 1$ the absolute value of  $\widetilde{\chi}^2$ gives a rough estimate of how
 inaccurate is the tested distribution to describe the data.

We fitted the data for the 5'UTR exons setting a minimum threshold on the exon length and then gradually increasing this threshold until a
reduced $\widetilde{\chi}^2$ value smaller than one was obtained. The rationale behind this choice is that (as we
shall see below)  the power law decay is likely to be an asymptotic behaviour which is violated for short exon
lengths. Starting from 
$l_{min}\sim 150$ both in human and in mouse good $\widetilde{\chi}^2$ values were obtained and we could
estimate the  critical
index to be  $\alpha\sim 2.5$. Detailed results of the fits are reported in Tab.1. The $\widetilde{\chi}^2$
values that we found support in a quantitative way
 the power law behaviour of the data, which was already evident
 looking at Fig.1a.\\

On the contrary, the coding exons and the 3'UTR exons length histograms display (on a ln-ln scale) non linear distributions with peaks of population around favourite sizes. In the range, where we are able to fit the power law decay of 5'UTR exons length, $\widetilde{\chi}^2$ values for linear fit in the other classes of exons are completely unacceptables (Tab. 2).

\begin{figure}[htbp]
\centering
\subbottom[\label{fig1a}]{\epsfig{file=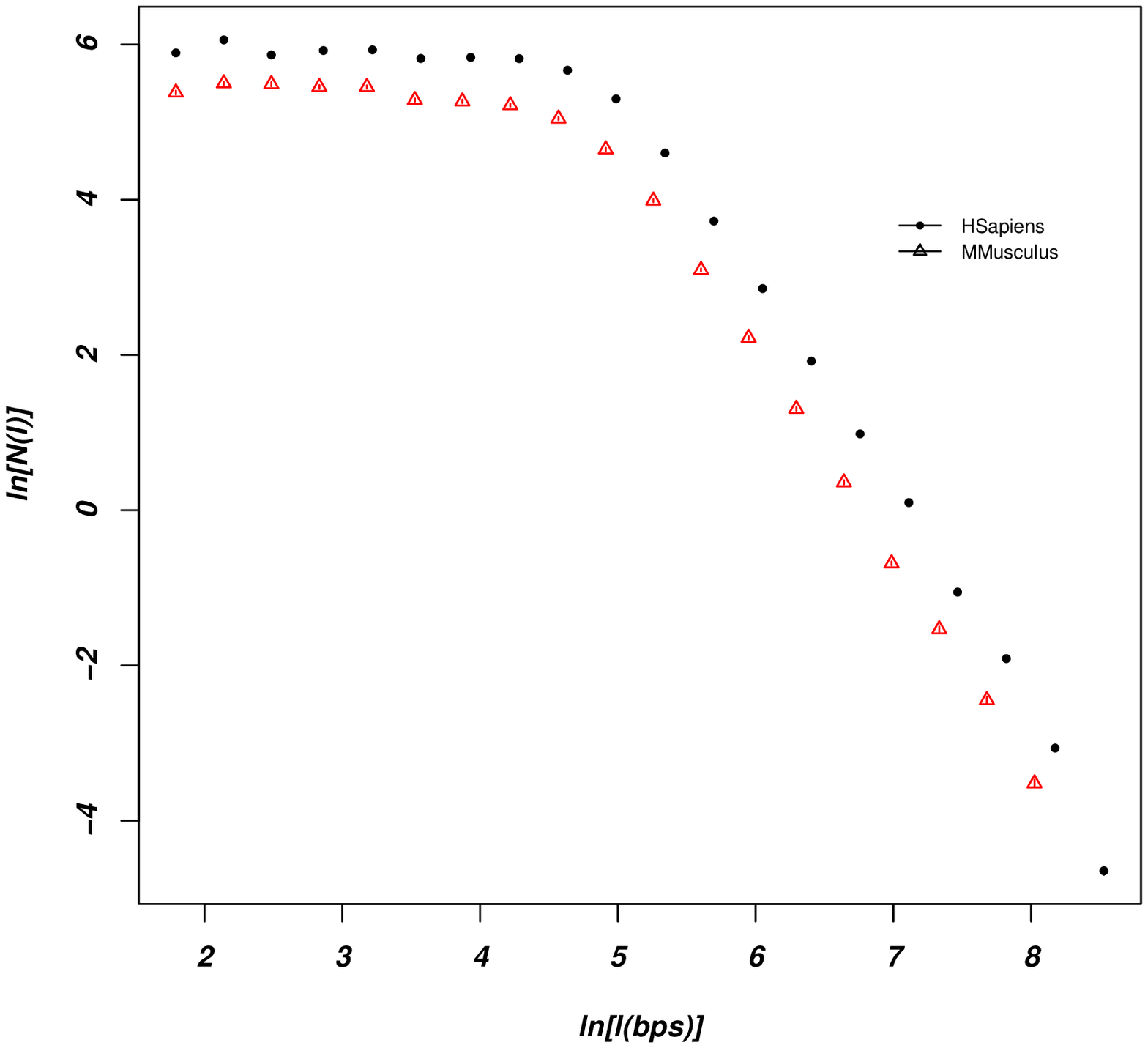,height=10cm,width=10cm,angle=-90}}
\subbottom[\label{fig1b}]{\epsfig{file=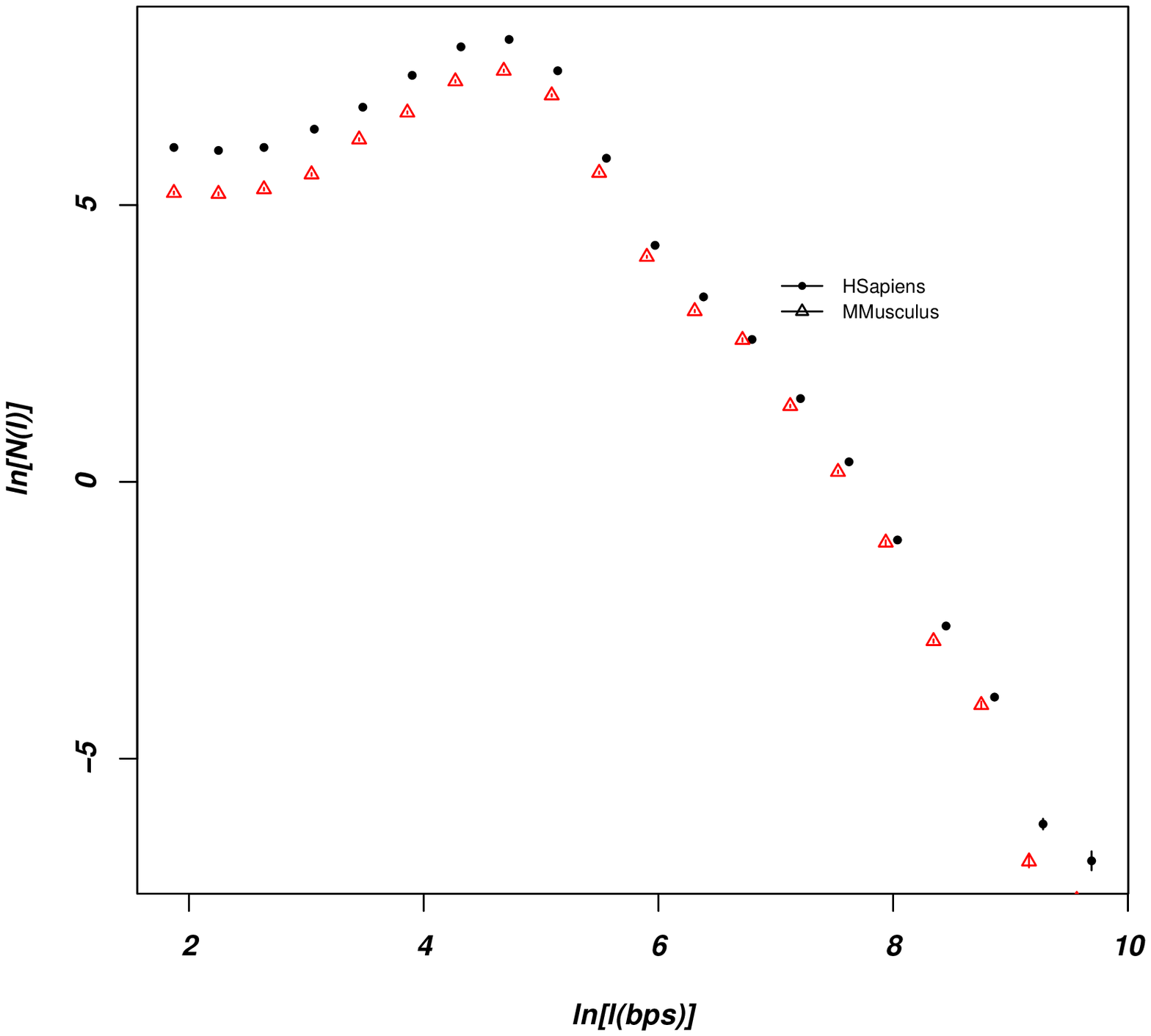,height=10cm,width=10cm,angle=-90}}
\end{figure}
\begin{figure}
\centering
\contsubbottom[\label{fig1c}]{\epsfig{file=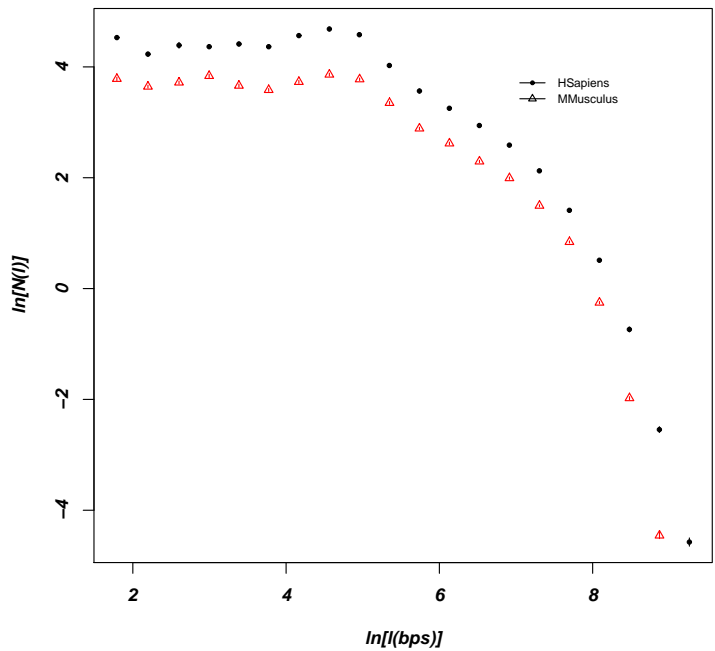,height=10cm,width=10cm,angle=-90}}
\caption{Exons length distribution in 5'UTR (a), protein coding exons (b) and 3'UTR (c) in human and mouse genome reported in ln-ln histograms (with bin size growing logaritmically). Plot errors are derived dividing the complete dataset in subsets of comparable dimension, avoiding biological biases, and averaging the length distribution of each subset.} 
\label{fig1}
\subconcluded
\end{figure}

\begin{table}[htbp]
\begin{center}
\begin{tabular}{|l|c|c|c|}
\hline
Species& $\widetilde{\chi}^2$ &$\alpha$ index& $l min$ (bps)\\
\hline
H.Sapiens&0.52&2.56(2)&150\\
M.Musculus&0.74&2.61(2)&140\\
\hline
\end{tabular}
\end{center}
\caption{Extimate of critical index $\alpha$ and length threshold $l min$ for the power law distribution of 
5'UTR exons in human and mouse}
\label{tab1}
\end{table}

\begin{table}[htbp]
\begin{center}
\begin{tabular}{|l|c|c|c|}
\hline
Species&protein coding exons&3'UTR exons& $l min$ (bps)\\
\hline
H.Sapiens&84.37&13.46&150\\
M.Musculus&153.31&5.91&140\\
\hline
\end{tabular}
\end{center}
\caption{ $\widetilde{\chi}^2$ values for the linear fit of protein coding exons and 3'UTR exons length distribution, in the same range where we are able to fit the power law decay of 5'UTR exons length}
\label{tab2}
\end{table}

The same plots for other organisms show exactly analogous trend, but they are affected by poor annotation of 
5' and 3' UTR, which are very difficult to identify entirely (see Tab.3). In Tab. 3 we reported the total number of annotated protein coding genes, annotated 5'UTR and annotated 3'UTR  for 4 different mammalian genomes, according to Ensemble database release 40. These data underline the current lack in the annotatation of 5'UTR and 3'UTR for other mammals, besides H. Sapiens and M. Musculus. For this reason, the same analysis performed for H. Sapiens and M. Musculus exon length distribution is prevented for other organisms.
\begin{table}[htbp]
\begin{center}
\begin{tabular}{|l|c|c|c|}
\hline
Species&Annotated protein coding genes&Annotated 5'UTR&Annotated 3'UTR\\
\hline
H.Sapiens&23735&18333&18592\\
M.Musculus&24438&15945&16429\\
C.Familiaris&18214&5925&6298\\
G.Gallus&18632&7463&7670\\
\hline
\end{tabular}
\caption{Annotated protein coding genes, 5'UTR and 3'UTR in Ensembl database release 40}
\label{tab3}
\end{center}
\end{table}
\vspace*{0.5 cm}

In order to understand this peculiar behaviour of the 5'UTR exons we propose and discuss in the following section a simple model of exon
evolution. Our goal is to understand if it is possible to associate the different behaviour that we observe
to the greater freedom from
selective pressure of the 5'UTR exons with
respect to the coding and 3'UTR ones.

\section{The model}
Evolutionary models describe evolution of the DNA sequence
as a series of stochastic mutations. There are three major classes of mutations: changes in the nucleotide type, 
insertions or deletions of one or more nucleotides. The various existing models differ with each other for the
different assumptions they make on the parameter which control these changes (for a review see for instance
\cite{durbin,mitchison,kosiol}). From a biological point of view
the two main assumptions of any evolutionary model are:
\begin{itemize}
\item 
evolution can be decribed as a Markov process, i.e. 
the modifications of a DNA sequence only depend on its current state and not 
on its previous history.
\item
evolution is ``shaped'' by functional constraints: DNA sequences with a negligible functional role evolve at a
higher rate with respect of functionally important regions. This implies that regions with different functional
roles must be described by different choices of the various mutational rates. The free evolution of sequences 
without functional constraints is usually called ``neutral evolution''. 
\end{itemize}
Let us see a few examples:
 \begin{itemize}
 \item
 protein coding exons are usually strongly constrained since the proteins they code have an important role in
 the life of the cell, however due to redundance of the genetic code, the third basis of each codon in the coding
 exons is free to mutate. On the contrary insertions and deletions are suppressed because they can dramatically
 affect the shape and function of the protein.
 \item
 Sequences devoted to transcriptional regulations
  (which very often lie outside exons) are usually so important for the life of the cell that
 they are kept almost unchanged over millions of years of evolution
 \item
 Regulatory sequences on the messanger RNA (mRNA) whose function often depends on the tridimensional shape of the
 RNA molecule and not on its exact sequence are in an intermediate situation between the above cases and the
 neutral evolution: they can tolerate mutations which
 do not modify their tridimensional shape (typically these are pairs of pointlike changes of bases 
 and are usually  called ``compensatory mutations''). 
 Most of the mRNA regulatory signals of this type are located in 3'UTR
 exons. 
 \item
 5'UTR regions contain sometimes regulatory sequences of the transcriptional type (which, as mentioned above, are
 stongly conserved under evolution) but their relative position seem not to have a crucial functional role. They
 can thus tolerate insertion and deletions as far as they do not affect the regulatory regions.
 \end{itemize}

Since in our model we are only interested in the exon length 
distribution we may neglect the nucleotide changes and
 concentrate only on insertions and deletions. From this point of view, according to the above discussion
 both coding and 3'UTR should behave as highly constrained sequences while the 5'UTR ones should be more similar
 to the neutrally evolving ones.
 With this picture in mind we decided to model the neutral evolution of a DNA sequence under the effect of
 insertions and deletions only, to see which general behaviour one should expect for the length distribution and
 then compare it with the data discussed in the previous section.

To this end let us define  $n_j$ as the number of 5' UTR exons of length $j$ in the genome and let $N$ be
the total number of such
exons. Let $x_j\equiv n_j/N$ be the fraction of exons of length $j$.

If we assume that the exon length distribution evolves as a consequence of insertions and/or deletions of single
 nucleotides we find the following evolution equation for the $x_j(t)$ (where $t$ labels the time step of this
 process)
 \eq
 x_j(t+1)=x_j(t) +(j-1)\alpha x_{j-1}(t)  -j\alpha x_{j}(t)+(j+1)\beta x_{j+1}(t) -j\beta x_{j}(t)
\label{eq} 
\en
 
where $\alpha$ and $\beta$ denote the insertion and deletion probabilities respectively and we have kept into account the fact that for an exon of length $j$ there are exactly $j$ sites in which the new nucleotide can be inserted (i.e. that the insertion and deletion probabilities are linear functions of $j$, since the implied assumption is that all sites in our sequences are independent of one another).

At equilibrium the exon length distribution must satisfy the following equation (we omit the $t$ dependence which
is now irrelevant)
 \eq
(j-1)\alpha x_{j-1}  -j\alpha x_{j}+(j+1)\beta x_{j+1} -j\beta x_{j}=0
\label{equi}
 \en

It is easy to see that the only solution compatible with this equation is a power law of this type:
$x_j=c j^{\eta}$ with $c$ a suitable
normalization constant. Inserting this proposal in eq.(\ref{equi}) one immediatly finds $\eta=-1$.

This result is very robust, it does not depend on the values of $\alpha$ and $\beta$ and, what is more important,
it holds also if instead of assuming the insertion (or deletion) of a single nucleotide, we assume the insertion
or deletion of oligos (i.e. small sequences of nucleotides) 
of length $k$, with any choice of the probability distribution for the oligos length as fas as $k$ is much
smaller than the typical exon length. Moreover one can also show that the power law decay still holds if we add
to the process a fixed background probability of creation of new exons of random length as far as this
probability is smaller than $x_{j_{max}}(\alpha-\beta)$ where $j_{max}$ is the largest exonic length for which
the power law is still observed. This is rather important since it is known that retrotransposed repeats (in
particular of the Alu family) may in some cases (with very low probability) 
become new active exons  and represent one of the major sources of evolutionary changes in the transcriptome.

On the contrary this power law disappears if we assume that there is a finite probability that,
as a consequence of the new insertion of deletion, the exon is eliminated. 
In this case the power law changes into a exponential distribution. This may explain why the power law decay is
not observed in the coding and 3'UTR portion of the genes which are under a much stronger selective pressure (in
the 3'UTR region are contained lot of post-trascriptional regulatory signals).

Since the critical index that we observe in the actual exon distribution in human and mouse is much larger than
$1$ it is interesting to see which type of evolutionary mechanism could lead to a $\eta>1$ behaviour while
keeping a power law decay. It is easy to see that this can be achieved assuming that the insertion (or deletion)
probability is not linear with the length of the exon but behaves, say, as $p_{insertion}=\alpha j^{\lambda}$
with $\lambda>1$. Then,  following the same derivation discussed above, we find at equilibrium an exon length
distribution $x_j=cj^{-\lambda}$. 

A possible explanation for such non-linear insertion rate comes from the observation that the transcribed
portions of the genome (like the 5' UTR exons in which we are interested), besides the normal mutation
processes typical of the intergenic regions, are subject to specific mutation events due to the transcriptional
machinery itself (see for instance~\cite{Touchon:2004}).

It is clear from the above discussion that in this case the critical index of the exon distribution, strictly
speaking, is not any more an universal
quantity, but depends on the particular  biological process  leading to the 
$p_{insertion}=\alpha j^{\lambda}$
probability discussed above. However it is conceivable that similar mechanisms should be at work in related
species. This in our opinion explains why the critical indices associated to the mouse and human 
distributions are so
similar and led us to conjecture that similar values should be found also in other mammalians as more and more
5'UTR sequences will be annotated.

Let us conclude by noticing that this whole derivation is based on the assumption that the system had reached its
equilibrium distribution. This is by no means an obvious assumption and it is well possible that the fact that we
observe a critical index larger than 1 simply denotes that the system is still slowly approaching 
the equilibrium distribution. There are three ways to address this issue. First one should extend 
the analysis to other
organisms (however, as we discussed above, 
this will require a better annotation of the UTR regions in these organisms). Second one could
reconstruct, by suitable aligning procedures, 
the UTR exons of the common ancestor between mouse and man and see if they also follow a power law distribution
and, if this is the case, which is the critical index. Third one could 
simulate the model discussed above and look
to the behaviour of the exon distribution as the equilibrium is approached. We plan to address these issue in a
forthcoming publication.

\vskip1.0cm {\bf Acknowledgements.} 
This work was partially supported by FIRB grant RBNE03B8KK from the Italian Ministry for
Education, University and Research.

The authors would like to thank D. Cor\`a, E. Curiotto, F. DiCunto, I. Molineris, P. Provero, A. Re and G. Sales for useful discussions and suggestions.

\appendix
\section{Derivation of the power law.}
Inserting the distribution $x_j=c j^{\eta}$ in eq.(\ref{equi}) we find 
 \eq
\alpha (j-1)^{\eta+1}  -\alpha (j)^{\eta+1}+\beta(j+1)^{\eta+1}  -\beta (j)^{\eta+1}=0
\label{app1}
 \en
which can be expanded in the large $j$ limit as 
 \eq
j^{\eta+1}\left[\alpha \left(1-\frac{\eta+1}{j}\right)  -\alpha +\beta \left(1-\frac{\eta-1}{j}\right)  -\beta
\right]=0
\label{app2}
 \en
which implies:
 \eq
(\beta-\alpha)\frac{\eta+1}{j}=0
\label{app3}
 \en
which (assuming $\beta\not=\alpha$) implies, as anticipated, $\eta=-1$.

A few observations are in order at this point:
\begin{description}
\item{a]} It is clear from the derivation that the result is independent from the specific values of $\alpha$ and
$\beta$ as far as they do not coincide. This independence from the details of the model holds 
also if we assume at
each time step a finite, constant (i.e. not proportional to $j$) 
probability $\alpha'~(\beta')$ of random insertion (deletion) of a nucleotide. In this case the evolution
equation becomes:
equation becomes:
 $$
 x_j(t+1)=x_j(t) +(j-1)\alpha x_{j-1}(t)  -j\alpha x_{j}(t)+(j+1)\beta x_{j+1}(t) 
 $$
 \eq
 -j\beta x_{j}(t)
+\alpha' (x_{j-1}(t)-x_j(t))+\beta'(x_{j+1}(t)-x_j(t))
\label{app4} 
\en
which still admits the same asymptotic distribution $x_j=c j^{-1}$
\item{b]}
If we include a fixed exonization probability $p_e$ to create new exons from, say, duplicated or
retrotransposed sequences the evolution equation changes trivially by simply adding such a constant contribution.
The solution becomes in this case $x_j=c j^{-1} +d$ where the constant $d$ is related to $p_e$ as follows
$d=p_e/(\alpha-\beta)$ and is negligible as far as it is smaller than $x_{j_{max}}$
\item{c]}
Remarkably enough the above results are still valid even if the inserted (or deleted) sequence is composed by
more than one nucleotide. Let us study as an example the situation in which we allow the insertion of oligos of
length $k$ with $0<k<L$ and L smaller than the typical exon length. Let us assume for simplicity to neglect
deletions and let us choose the same
insertion probability $\alpha$ for all values of $k$. The evolution equation becomes:
 \eq
 x_j(t+1)=x_j(t) +\alpha \left[\sum_{k=1}^{L}x_{j-k}(t)(j-k)  - L jx_{j}(t)\right]
\label{app5} 
\en
which implies
 \eq
jx_{j}=\frac1L \sum_{k=1}^{L}(j-k)x_{j-k} 
\label{app6} 
\en

In the large $j$ limit this equation admits again a power law solution $x_j=c j^{\eta}$. Inserting this solution
in eq.(\ref{app5}) we find
 \eq
j^{\eta+1}\alpha \left[\frac1L \sum_{k=1}^{L}\left(1-\frac{k(\eta+1)}{j}\right)  -1\right]=0
\label{app7}
\en
which is satisfied, as above,  if we set $\eta=-1$.
\item{d]}
On the contrary, if we assume a finite probability $(1-\gamma)$ of elimination of an exon as a consequnce of the
insertion (or deletion) event (as one would expect if the sequence is under strong selective pressure) we find
the following evolution equation:
 \eq
 x_j(t+1)=x_j(t) +[(j-1)\alpha x_{j-1}(t)\gamma -j\alpha x_{j}(t)]
\label{app8} 
\en
where $\alpha$ is, as above, the insertion probability and we are assuming for simplicity single base insertions.
This equation does not admit any more a power law solution at equilibrium but requires an exponential
distribution: $x_j=e^{-\lambda j}j^{\eta}$ with $\eta=-1$ and $\lambda=ln(\gamma)$.
\item{e]}
It is instructive to reobtain the result discussed in [a] above by looking at the equilibrium equation as a
recursive equation in $j$:
\eq
x_{j+1}=\frac{j}{j+1}\left(1+\frac{\alpha}{\beta}\right)x_j-\frac{\alpha}{\beta}x_{j-1}~~~~ (j>j_{min})
\en
and
\eq
x_{j+1}=\frac{j}{j+1}(1+\frac{\alpha}{\beta})x_j ~~~~(j=j_{min})
\en
and 
construct recursively the solution for any $j$ starting from $x_{j_{min}}=c/j_{min}$. The recursion can be solved
exactly and gives:
\eq
x_j= x_{j_{min}}\frac{j_{min}}{j}\frac{1-\left(\frac{\alpha}{\beta}\right)^{j-j_{min}+1}}{1-\frac{\alpha}{\beta}}
\en
which (assuming $\alpha<\beta$)~\footnote{If $\beta<\alpha$ one should study the inverse recursion relation
starting from $x_{j_{max}}$.} leads asymptotically to the solution $x_j=c/j$ with 
$c=x_{j_{min}}\frac{j_{min}}{1-\alpha/\beta}$. 
This result allows to understand exactly the ``finite size'' corrections with
respect to this asymptotic solution which turn out to be proportional to 
$\left(\frac{\alpha}{\beta}\right)^{j-j_{min}+1}$ and vanish if only deletions (i.e. $\alpha=0$) 
or only insertions
(i.e. $\beta=0$) are present. In these cases the asymptotic solution is actually the {\sl exact} equilibrium
solution of the stochastic model.
 
\end{description}

\end{document}